\documentclass[fleqn,usenatbib]{mnras}

\usepackage{newtxtext,newtxmath}

\usepackage[T1]{fontenc}

\DeclareRobustCommand{\VAN}[3]{#2}
\let\VANthebibliography\thebibliography
\def\thebibliography{\DeclareRobustCommand{\VAN}[3]{##3}\VANthebibliography}

\usepackage{graphicx}	
\usepackage{amsmath}	

\title[The current sheet formation a coronal jet event]{Simultaneous observations of a breakout current sheet and a flare current sheet in a coronal jet event}

\author[L. H. Yang et al.]{
Liheng Yang,$^{1,2}$\thanks{E-mail: yangliheng@ynao.ac.cn}
Xiaoli Yan,$^{1,2}$
Zhike Xue$^{1,2}$
Zhe Xu$^{1,2}$
Qingmin Zhang$^{3}$
Yijun Hou$^{4}$
Jincheng Wang$^{1,2}$
\newauthor{Huadong Chen$^{4}$ and Qiaoling Li$^{5}$}
\\
$^{1}$Yunnan Observatories, Chinese Academy of Sciences, Kunming 650216, China\\
$^{2}$Yunnan Key Laboratory of the Solar physics and Space Science,Kunming 650216\\
$^{3}$Purple Mountain Observatory, Chinese Academy of Sciences, Nanjing 210023, China\\
$^{4}$National Astronomical Observatories, Chinese Academy of Sciences, Beijing 100012, China\\
$^{5}$Department of Physics, Yunnan University, Kunming, Yunnan 650091, China\\
}

\date{Accepted XXX. Received YYY; in original form ZZZ}

\pubyear{2015}

\begin{document}
\label{firstpage}
\pagerange{\pageref{firstpage}--\pageref{lastpage}}
\maketitle

\begin{abstract}
Previous studies have revealed that solar coronal jets triggered by the eruption of mini-filaments (MFs) conform to the famous magnetic-breakout mechanism. In such scenario, a breakout current sheet (BCS) and a flare current sheet (FCS) should be observed during the jets. With high spatial and temporal resolution data from the SDO, the NVST, the RHESSI, the Wind, and the GOES, we present observational evidence of a BCS and a FCS formation during coronal jets driven by a MF eruption occurring in the active region NOAA 11726 on 2013 April 21. Magnetic field extrapolation show that the MF was enclosed by a fan-spine magnetic structure. The MF was activated by flux cancellation under it, and then slowly rose. A BCS formed when the magnetic fields wrapping the MF squeezed to antidirectional external open fields. Simultaneously, one thin bright jet and two bidirectional jet-like structures were observed. As the MF erupted as a blowout jet, a FCS was formed when the two distended legs inside the MF field came together. One end of the FCS connected the post-flare loops. The BCS’s peak temperature  was calculated to be 2.5 MK. The FCS’s length, width and peak temperature  was calculated to be 4.35-4.93 Mm, 1.31-1.45 Mm, and 2.5 MK, respectively. The magnetic reconnection rate associated with the FCS was estimated to be from 0.266 to 0.333. This event also related to a type III radio burst, indicating its influence on interplanetary space. These observations support the scenario of the breakout model as the trigger mechanism of coronal jets, and flux cancellation was the driver of this event.
\end{abstract}




\section{Introduction}

Solar jets are small-scale transient eruptions observed as apparent collimated flows of plasma \citep{1992PASJ...44L.173S}. They are prevalent in active regions (ARs), coronal holes, and quiet sun regions. These small-scale jets might contribute to coronal heating and solar wind acceleration, because they can continuously supply mass and energy into the upper solar atmosphere \citep{1997Natur.386..811I,2007Sci...318.1591S,2014Sci...346A.315T}. Solar jets bear a strong resemblance to large-scale explosive events, such as flares, filament eruptions, and coronal mass ejections (CMEs), especially in the aspect of the driving mechanisms \citep{2016SSRv..201....1R}. Solar jets can be detected in white light, H$\alpha$, Ca \small{II} H, EUV, and soft X-ray channels, and usually exhibit helical structures and untwisting motions \citep[e.g.,][]{1973SoPh...28...95R,1995SoPh..156..245S,2011RAA....11.1229Y,2012ApJ...745..164S,2019ApJ...887..154L,2019FrASS...6...44L,2023arXiv230914871M}. In a jet event, hot and cool plasma ejections are usually observed successively. The cool jet component might come from the eruptive mini-filament (MF) of the jet base, while the hot component might be the plasma heated by magnetic reconnection  \citep{2017ApJ...851...67S}. Sometimes, coronal jets are confined by the AR loops, and such kind of jets might be the result of magnetic reconnection taking place between MFs and its overlying large-scale loops \citep{2019ApJ...887..239Y}. Coronal jets can also drive EUV waves \citep[e.g.,][]{2017ApJ...851..101S,2019ApJ...871..232Z,2022ApJ...931..162Z}. Several reviews have summarized the observational characteristics and related theoretical models of solar jets \citep{2016AN....337.1024I,2016SSRv..201....1R,2021RSPSA.47700217S}.

MFs are initially observed as small fibrils or filament-like structures, which would expand into arches, break open at their top, and disappear \citep{1986NASCP2442..369H}. They have a typical projected lengths of 19 Mm, an ejection speed of 13 km s$^{-1}$, and a mean lifetime of 50 minutes \citep{2000ApJ...530.1071W}. They usually lie along the magnetic neutral line, and are related to cancelling magnetic features, which are considered to be the small-scale analog to large-scale filaments \citep{2020ApJ...902....8C}. Recent observations reveal that most solar jets are accompanied by the eruptions of MFs \citep{2010ApJ...720..757M,2012ApJ...745..164S,2015ApJ...814L..13L,2015Natur.523..437S,2016ApJ...821..100S, 2016ApJ...830...60H,2017ApJ...835...35H,2019ApJ...887..239Y,2019ApJ...883..104S,2023ApJ...945...96Y}. Such kind of solar jets are defined as blowout jets by \citet{2010ApJ...720..757M} on the basis of the structure and development of X-ray jets. Another type they defined is standard jet, which is generated by the magnetic reconnection between the emerging closed field and surrounding open field \citep{1992PASJ...44L.173S,1995Natur.375...42Y}. The blowout jet's base arch contains a filament or flux rope, which is generally sheared and twisted. That is the largest difference between the blowout and standard jets. According to further observations of \citet{2015Natur.523..437S}, the formation of almost all the coronal jets is associated with erupting MFs. Whether the MF erupts successfully or unsuccessfully determines that the jet is a blowout or a standard type.

Many observations and numerical simulations show that coronal jets are associated with magnetic reconnection occurring in a fan-spine magnetic topology \citep{2009ApJ...704..485T,2011ApJ...728..103L,2012A&A...545A..78W,2016ApJ...827...27Z,2019ApJ...885L..11S,2020ApJ...900..158Y,2022ApJ...926L..39D}. A fan-spine magnetic topology usually appears if a magnetic bipole emerges into a unipolar magnetic field. It is made up of a coronal null point, a outer spine, a inner spine, and a dome-like fan. The dome-like fan represents the closed separatrix surface dividing two different connectivity domains, while the inner and outer spines belong to different connectivity domains \citep{1990ApJ...350..672L,2021RSPSA.47700217S}. When a magnetic bipole continuously emerges within the central parasitic region encompassed by the opposite-polarity fields, the secondary fan-spine structure would form under the large fan \citep{2019ApJ...871....4H}. Recent observations show that eruption of MFs or sheared flux-rope structures under the fan dome can result in null point reconnection and the formation of solar jets \citep{2017ApJ...842L..20L,2019ApJ...872...87L,2016ApJ...827...27Z,2021A&A...647A.113Z}. A breakout model for solar coronal jets involving MF eruptions in a fan-spine structure was simulated by \citet{2017Natur.544..452W} and \citet{2018ApJ...852...98W}. Their model is an extension of the CME breakout model. To model this process, a potential field on the solar surface was adopted, in which a compact bipolar magnetic structure was embedded in a uniform inclined background field. As the footpoint driving proceeded, the sheared filament channel with a MF formed and expanded near the center of the bipole field. With the increase of the magnetic pressure within the bipole region, the null point above the bipole region became compressed, and a breakout current sheet (BCS) formed there. Subsequently, reconnection removed some of the strapping field above the MF, and the MF began to lift upward. When it reached the BCS, the magnetic field carrying the MF (or a flux rope) reconnected with the external open field, forming the jet spire. The continuous upward lifting of the flux rope formed a flare current sheet (FCS) below the MF. The FCS connected the flux rope and the jet bright point. 

Magnetic reconnection is a basic process in magnetized plasma, providing an efficient method for conversion magnetic energy into the heat, nonthermal energy, and kinetic 
energy \citep[e.g.,][]{2000mare.book.....P,2009A&A...508.1443B,2011ApJ...742...82K,2012ApJ...754....9G,2015SSRv..194..237L,2018ApJ...867...84G,2022LRSP...19....1P}. It takes responsibility for solar flares, jets, CMEs, the solar wind, coronal heating, etc. When oppositely directed magnetic fields are severely stretched, a strong electric current region generally appears in the form of a current sheet (CS). CS is one of significant topological structures of magnetic reconnection. It is a thin diffusion region allowing fast magnetic flux transfer. Large-scale CSs have been widely observed in CME-flare events \citep[e.g.,][]{2003ApJ...594.1068K,2007ApJ...658L.123L,2011ApJ...742...92W,2018ApJ...866...64C,2018ApJ...853L..18Y}. Thanks to high-resolution observations, small-scale CSs have been detected in the reconnection region between small-scale loops \citep{2015ApJ...798L..11Y}, filament and chromospheric fibrils \citep{2016NatCo...711837X}, filament and loops \citep{2016NatPh..12..847L}, filament and emerging field \citep{2022ApJ...935...85L}, twisted flux rope surrounding a filament and magnetic loops \citep{2022NatCo..13..640Y}. These small-scale CSs have also been observed in coronal jets. An equatorial coronal-hole jet associated with a MF eruption was studied by \citet{2018ApJ...854..155K}, and they clearly observed the BCS and the FCS during the eruption process. Their observation matched the predictions of the breakout jet model. \citet{2019ApJ...879...74C} studied the CS formation in two quiet-Sun jets associated with MFs. They found that the length, width, and temperature of the CS were around 7.51-10.8 Mm, 1.86-3.4 Mm, and 1.8 MK, respectively. \citet{2019ApJ...874..146H} investigated a jet produced by the eruption of two MFs, and found the oscillatory magnetic reconnection of the BCS. Recently, \citet{2023ApJ...942...86Y} observed weak bidirectional outflows and a FCS in a coronal-hole jet driven by the eruption of a MF. 

In this paper, we study coronal jets associated with a MF ejecting from the northern edge of the following spot in the AR NOAA 11726 with high-resolution, multi-wavelength observations from the Solar Dynamics Observatory \citep[SDO;][]{2012SoPh..275....3P} and the New Vacuum Solar Telescope \citep[NVST;][]{2014RAA....14..705L,2020ScChE..63.1656Y}. This event perfectly reproduced the two key physical processes of the breakout jet model simulated by \citet{2018ApJ...852...98W}. The BCS and the FCS were clearly observed in the eruption process. We introduce the observational data in Section 2, show the results in Section 3, and present the conclusions and discussions in Section 4.

\section{Observations and Data Analysis}

The one-meter NVST aims to detect small-scale structures of the Sun's photosphere and chromosphere. Its main observational equipments are a multi-channel high resolution imaging system and a multi-band spectrometer, which are installed on a 6-meter rotating platform and rotate along with the platform. The multi-channel high resolution imaging system can provide images in one photosphere channel (TiO 7058~{\AA}) and two chromosphere channels (H$\alpha$ 6563~{\AA}  and He I 10830~{\AA}). A tunable Lyot filter is used in the H$\alpha$ channel. The filter is centered at 6562.8~{\AA}, and can scan spectra in the range of $\pm$5~{\AA} with a step size of 0.1~{\AA}. It has a bandwidth of 0.25~{\AA}.
A high-order solar adaptive optics system (AO) has been installed at NVST since 2016 \citep{2016ApJ...833..210R,2016SPIE.9909E..2CZ}. It is made up of a high order wavefront correction loop and a fine tracking loop. The main component of the system 
are a deformable mirror, a correction Shack-Hartmann wavefront sensor, and a custom-built real-time controller. The deformable mirror contains 151 actuators, and the correction Shack-Hartmann wavefront sensor consists of 102 sub-apertures. Based on 
a Field-Programmable Gate Array (FPGA) and multi-core Digital Signal Processor (DSP), the custom-built real-time controller is set up. In 2021, a new multi-direction wavefront sensor was added to achieve the GLAO correction mode, the hardware platform was updated with FPGA+CPUs architecture to meet the real-time processing and controlling requirement of both AO and GLAO system \citep{2023SCPMA..6669611Z}. In the present work, only H$\alpha$ line center images were used. The pixel size and temporal resolution of these images are 0.165 $''$ and 12~s, respectively. The raw H$\alpha$ data need to be subtracted the dark current, corrected the flat field, and then reconstructed by using the speckle masking method \citep{1977OptCo..21...55W,1983ApOpt..22.4028L,2016NewA...49....8X}.
 
The Atmospheric Imaging Assembly \citep[AIA;][]{2012SoPh..275...17L} on board SDO consists of four generalized Cassegrain telescopes. It can provide multi-wavelength, high-resolution full-disk images from transition region up to corona. These images have a pixel size of 0.6 $''$ and a temporal resolution of 12~s. They are obtained in seven EUV passbands centered on 94~{\AA}, 131~{\AA}, 335~{\AA}, 211~{\AA}, 193~{\AA}, 171~{\AA}, and 304~{\AA} and three continuum bands centered on 4500~{\AA}, 1700~{\AA}, and 1600~{\AA}. Their temperature diagnostics cover the range from 0.05 to 20 MK. By using the almost simultaneous observations of six AIA EUV lines centered on 131~{\AA}, 94~{\AA}, 335~{\AA}, 211~{\AA}, 193~{\AA}, and 171~{\AA}, the differential emission measure (DEM) can be reconstructed \citep[e.g.,][]{2012ApJ...761...62C}. In this event, the emission measures are calculated and used. The Helioseismic and Magnetic Imager (HMI) on board SDO can provide four type of the data, including line-of-sight magnetograms, continuum filtergrams, dopplergrams, and vector magnetograms. The line-of-sight magnetograms are taken every 45 seconds, and have a 1 $''$ resolution. The vector magnetograms have the same spatial resolution with the line-of-sight magnetograms, but are taken every 12 minutes. On the basis of a nonlinear force-free field (NLFFF) method, we obtain the coronal magnetic field by extrapolating the vector magnetic field maps \citep{2004SoPh..219...87W}. However, the photospheric magnetic field is not preprocessed before the extrapolation. According to \citet{2017ApJ...839...30F}, though the preprocessing can meet the force-free boundary condition, it can introduce a systematic error in the height scale of extrapolating field. Recently, \citet{2023A&A...679A...9B} extrapolated the braiding loops by using the NLFFF method with no preprocessing, and the extrapolated result well agreed with the observations. The AIA and HMI line-of-sight data need to be processed by using 
the standard routine aia$\_$prep.pro in the SolarSoftWare (SSW) packages \citep{1998SoPh..182..497F}. 

The Reuven Ramaty High-Energy Solar Spectroscopic Imager (RHESSI) is designed to investigate particle acceleration and energy release in solar flares, through 
imaging and spectroscopy of hard X-ray/gamma-ray continua (3 keV) emitted by energetic electrons, and of gamma-ray lines (17 MeV) produced by energetic ions \citep{2002SoPh..210....3L}. 
The spatial resolution is $\sim$2.3 $''$, and the spectral resolution is $\sim$1-10 keV. In this paper, the RHESSI data are used to study the hard X-ray sources during the B-class microflare, which 
is constructed by the clean imaging algorithm. The accumulation time is 1 minute. We selected the nonthermal energy range of 12-25 keV.

The WAVES experiment on the Wind spacecraft provides measurements of the radio and plasma wave phenomena occurring in the frequency ranging from a fraction of Hertz up about 14 MHz for the 
electric field and 3 kHz for the magnetic field \citep{1995SSRv...71..231B}. It consists of three orthogonal search coil magnetometers and three orthogonal electric field antenna. Five different receivers are used 
to measure the electric fields, including thermal noise receiver (TNR, 4-256 kHz), low frequency fast Fourier transform (FFT) receiver (0.3 Hz to 11 kHz), the time domain sampler (TDS), 
radio receiver band 1 (RAD1, 20-1040 kHz), and  radio receiver band 2 (RAD2, 1.075-13.825 MHz). In this paper, the RAD1 and the RAD2 data were used.

The X-ray sensor (XRS) onboard the Geostationary Operational Environmental Satellite (GOES) provides solar X-ray fluxes for the wavelength bands of 0.5-4~{\AA} and 1-8~{\AA}. In this event, the 
GOES 1-minute averages of soft X-ray in 1-8~{\AA} was used.

\section{Results}

\subsection{The Source Region of Coronal Jets}

On 2013 April 21, two coronal jets ejected from the northern edge of the AR NOAA 11726. This AR is located at N13$^\circ$W19$^\circ$, which has a $\beta$$\gamma$-type. The coronal jets were triggered by the 
eruption of a MF. Figure 1 shows the location of the MF and its surrounding magnetic environment before eruption. From the HMI line-of-sight magnetogram, it is noted that the jet base has a characteristic that a central negative polarity (N) is surrounded by the positive polarities (P) (see Figure 1(a)). Such type of the magnetic field is easy to form a fan-spine topology. The simultaneous SDO/AIA 304~{\AA} image reveals that the jet base has a bright anemone-like structure (marked by the white arrow in Figure 1(b)). A slightly curved spire was above the bright anemone-like structure, and seemed to be around for a long time (see the animation 1). The bright anemone-like structure might be the dome of the fan-spine topology, while the spire might be the outer spine. Generally speaking, the anemone-like base can produce anemone jets \citep{1994ApJ...431L..51S}. It is noted that the MF is situated on the west of the bright dome, which exhibits a reverse-S shaped dark structure. The reverse-S shaped MF in the north accords with the hemispheric pattern of magnetic helicity \citep{1998SoPh..182..107M}. Figures 1(c)-(e) present a zoomed-in HMI line-of-sight magnetogram, NVST H$\alpha$ image, and SDO/AIA 193~{\AA} image, which can show the general appearance of the MF and its underlying magnetic field in more detail. The MF outline (see the blue curve) depicted from the H$\alpha$ image at $\sim$06:20 UT is superimposed on Figure 1(c). The field of view of these images is given by the gold square in Figure 1(a). It is clear that the MF lies on the polarity inversion line (PIL), whose north (south) end rooted in the negative (positive) magnetic field region. The axial field of the MF is directed northward. According to the definition provided by \citet{1994ASIC..433..303M}, the MF is dextral. It should have a negative helicity with left-handed chirality \citep{2000ApJ...540L.115C}. The MF appears as a thin dark structure in the NVST H$\alpha$, SDO/AIA 304~{\AA}, and 193~{\AA} images. Its projected length is about 17 Mm, corresponding to the average length of MFs given by \citet{2000ApJ...530.1071W}. 

\begin{figure*}
       \centering
	\includegraphics[width=0.8\textwidth]{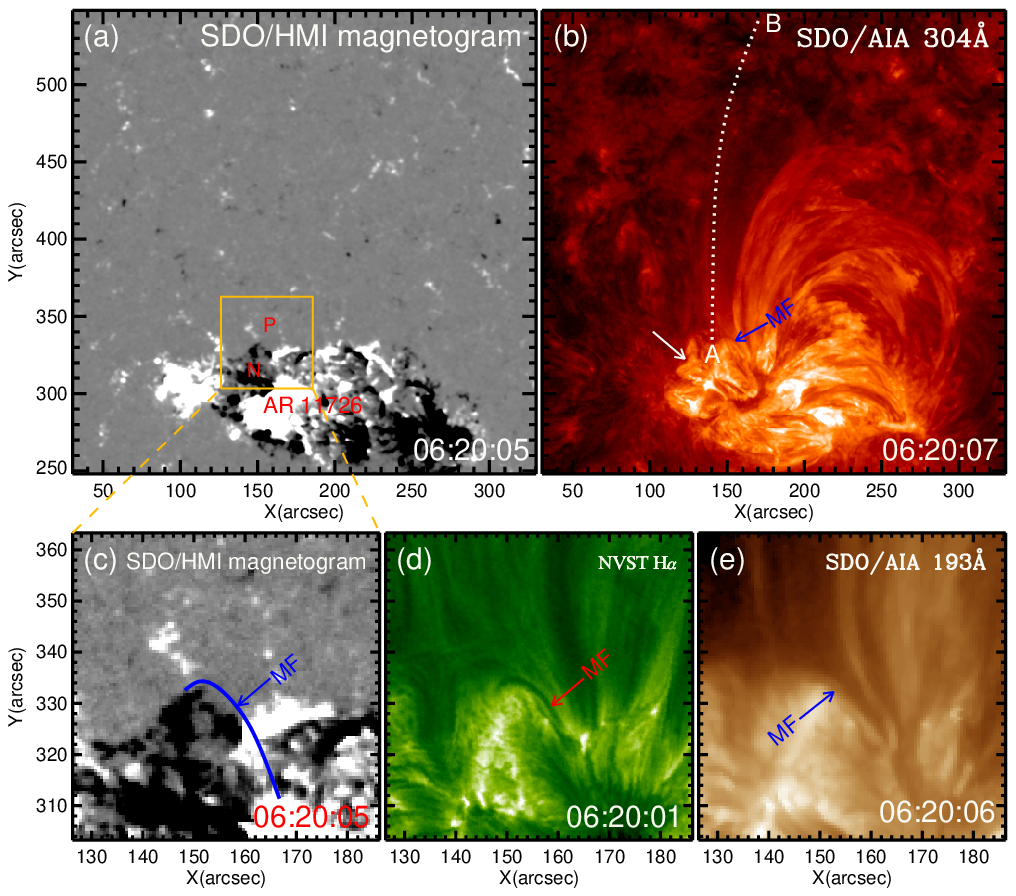}
    \caption{(a) HMI line-of-sight magnetogram and (b) SDO/AIA 304~{\AA} image showing the location of the jet. (c) HMI line-of-sight magnetogram, (d) NVST H$\alpha$ image, and (e) SDO/AIA 193~{\AA} image showing the mini-filament (MF) associated with the jet. The blue line in panels (c) denotes the MF derived from the H$\alpha$ image in panel (d). The gold square in panel (a) gives the field of view (FOV) of panels (c -- e). The slit position of the space-time plots in Figure 6 is marked by the curve line from ``A'' to ``B''.}
    \label{fig:fig1}
\end{figure*}

\begin{figure*}
	\includegraphics[width=0.8\textwidth]{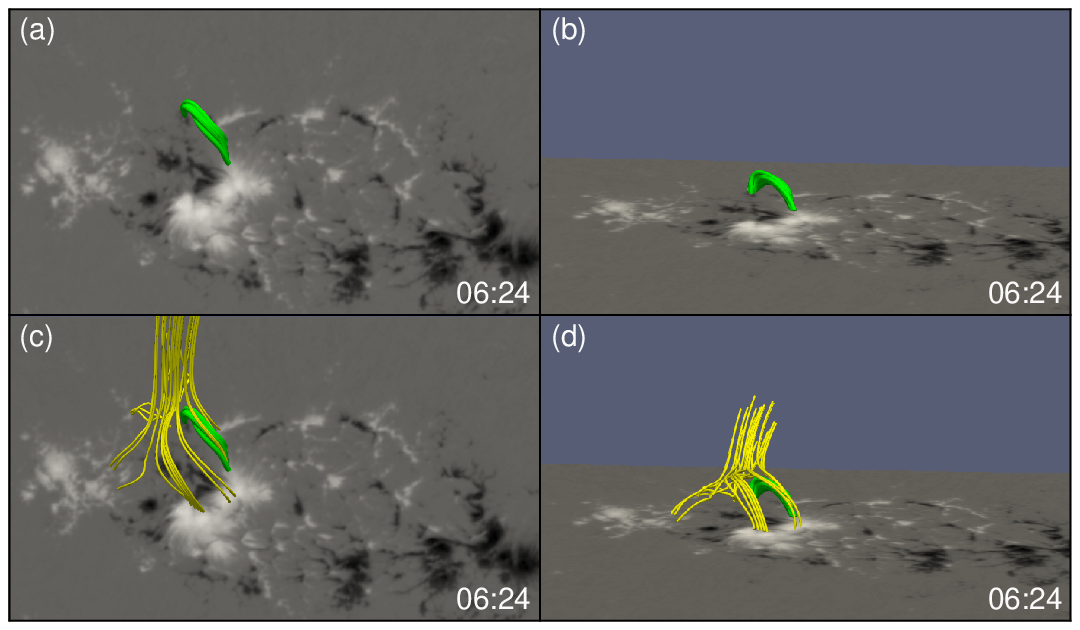}
    \caption{The NLFFF magnetic topology for the jet source region from top (a and c) and side (b and d) view. The green lines represent the MF. The yellow lines represent the fan-spine structure above the MF.}
    \label{fig:fig2}
\end{figure*}

Figure 2 displays the NLFFF magnetic topology of the source region from the top (Figures 2(a) and (c)) and side views (Figures 2(b) and (d)) before the eruption. As shown in Figures 2(a) and (b), a series of twisted magnetic arcades (green lines) just lies on the PIL of the source region. It connects the parasitical negative magnetic polarities and the dominant positive polarities. It might be the magnetic structure of the MF. The existence of the MF suggests that non-potential magnetic energy has been stored inside the fan. In Figures 2(c) and (d), it is found that a fan-spine structure (yellow lines) is just above the MF, and it is rooted in the surrounding positive polarities. The MF resides on the west of dome. Such extrapolated scenario is consistent with the observations.

\subsection{The Activation of the MF and the Formation of the BCS}

\begin{figure*}
	\includegraphics[width=0.9\textwidth]{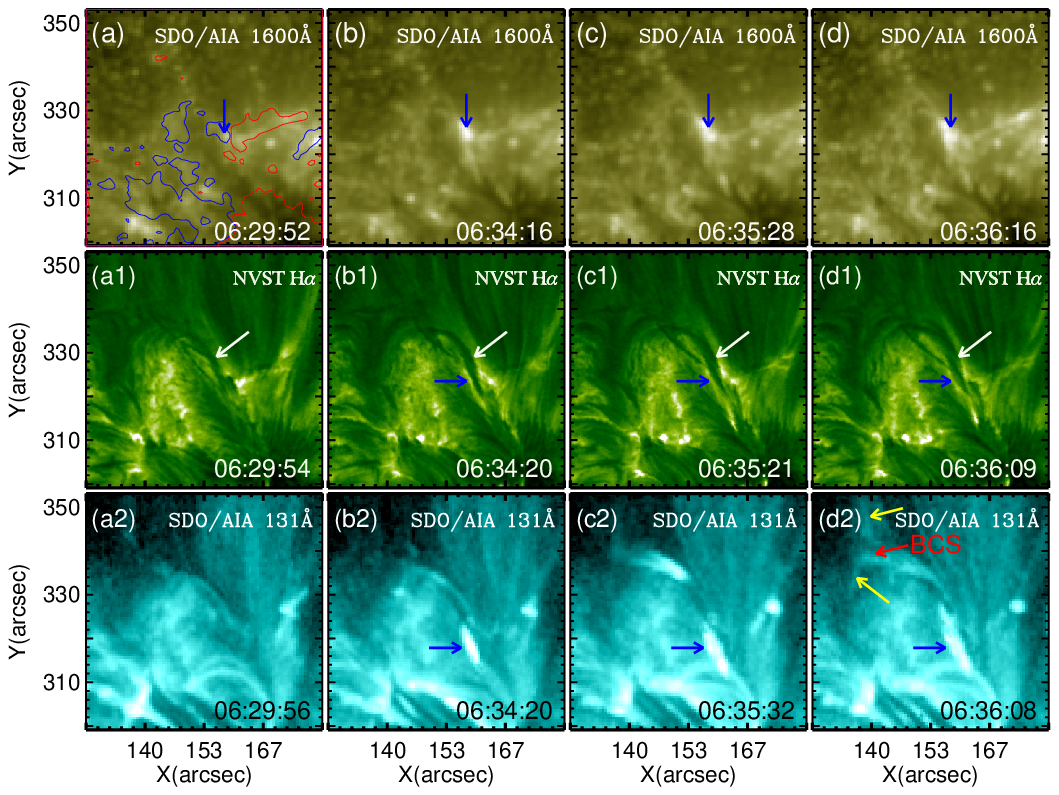}
    \caption{(a -- d): SDO/AIA 1600~{\AA}, (a1 -- d1): NVST H$\alpha$, and (a2 -- d2): SDO/AIA 131~{\AA} images showing the activation of the MF and the formation of the breakout current sheet (BCS). Blue/red contours represent negative/positive polarity regions, and the contour levels are $\pm$150 Gauss. The blue arrows point to the brightenings associated with the MF. The white arrows point to the activated MF. The yellow arrows point to bidirectional material flows. The red arrow points to the breakout current sheet (BCS).}
    \label{fig:fig3}
\end{figure*}

Figure 3 presents the activation of the MF and the formation of the BCS in SDO/AIA 1600~{\AA}, NVST H$\alpha$, and SDO/AIA 131~{\AA} images, respectively. The almost simultaneous SDO/HMI line-of-sight magnetogram is superimposed on the 1600~{\AA} image (see Figure 3(a)). The blue and red contours represent the negative and the positive magnetic field with levels of $\pm$150 Gauss. At $\sim$06:29 UT, a brightening appeared at the jet base. It manifested as a bright patch in the 1600~{\AA} wavelength (see the animation 2). However, no corresponding signals were detected in H$\alpha$ and 131~{\AA} wavelengths (see Figures 3(a1) and (a2)). As shown in Figure 3(a), the brightening (marked by the blue arrow) was located on the contact position of the negative and the positive magnetic field, where magnetic cancellation might occur. As time went on, the brightening observed in the 1600~{\AA} wavelength got bigger and brighter (see the animation 2). At $\sim$06:33 UT, a brightening started to appear in H$\alpha$ wavelength. Simultaneously, the brightening in the 1600~{\AA} wavelength began to spread northward and southward. About half a minute later, a brightening began to appear in the 131~{\AA} wavelength (see Figure 3(b2)). There exists a time lag for the appearance of the brightening in different wavelengths, indicating that magnetic reconnection might occur in low corona. It is noted that the brightening in the 1600~{\AA} and 131~{\AA} wavelengths rapidly spread along the MF after appeared. At $\sim$06:34 UT, the MF started to be activated. It obviously became thicker and darker (see Figure 3(c1)). Subsequently, the MF began to rise slowly (see the animation 2). At $\sim$06:35 UT, the magnetic field carrying the MF seemed to collide with the external open field. Later, the CS like structure with enhanced emission formed at the collision site (see Figure 3(d2)). At the same time, bi-directional jet-like structures (marked by yellow arrows in Figure 3(d2)) appeared to move along the spine. One of them was along the outer spine lines, and the other one was along the inner spine lines. These observations reveal that breakout reconnection might occur at the collision site, where the null point of the fan-spine structure might situate. The CS like structure might be the BCS. The BCS existed for 2 minutes, and was not observed in 1600~{\AA} and H$\alpha$ wavelengths. Normally, a sheet-like structure and two bifurcated structures at the two end points of the sheet-like structures should be observed in the magnetic reconnection process, like \citet{2019ApJ...874..146H}. Here, only one bifurcated structure was observed, perhaps due to the observation angle. 

\subsection{The Magnetic Field Evolution of the Source Region in the photosphere}

\begin{figure*}
	\includegraphics[width=0.9\textwidth]{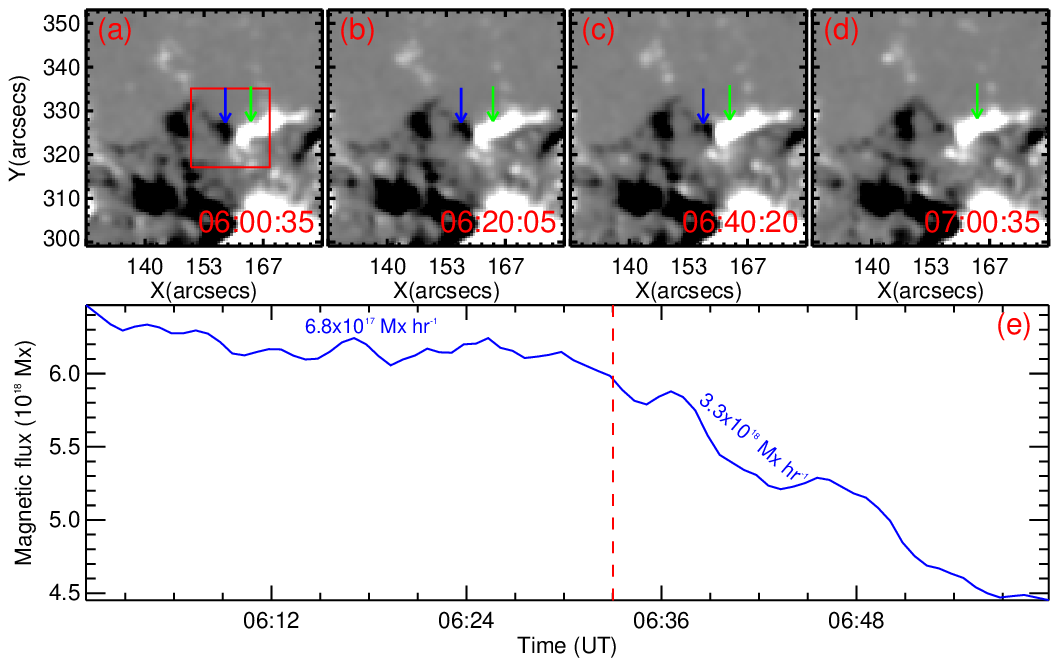}
    \caption{(a -- d): HMI line-of-sight magnetograms displaying the evolution of the magnetic field under the MF. (e): The negative magnetic fluxes changes in the red box region, as shown in panel (a). The blue/red arrows point to the location of the parasitical negative/the dominated positive polarities associated with the magnetic cancellation. The red dashed line represents the start time of the event.}
    \label{fig:fig4}
\end{figure*}

As mentioned earlier, the brightening appeared on the contact position of the negative and the positive magnetic field, which indicates that the photospheric magnetic field might lead to the eruption of the MF and the subsequent coronal jets. Figure 4 shows the evolution of the HMI photospheric magnetic field in the source region. The positive and the negative magnetic polarities were represented by the white and the black patches, respectively. The data adopted here are half an hour before and after the eruption onset. At $\sim$06:00 UT, the parasitical negative (marked by the blue arrows in Figures 4(a)-(c)) and the dominated positive magnetic polarities (marked by the green arrows in Figures 4(a)-(d)) under the MF came together (see Figure 4(a)). As time went on, the area of the negative magnetic polarities became smaller and smaller (see the animation 3). At $\sim$07:00 UT, the negative magnetic polarities under the MF almost completely disappeared. This evolution suggests that magnetic cancellation occurred before and after the MF eruption. Previous observations reveal that magnetic cancellation under MFs can drive the eruption of the MFs \citep[e.g.,][]{2016ApJ...832L...7P,2018ApJ...853..189P,2019ApJ...879...74C,2018ApJ...864...68S,2017ApJ...844...28S,2019ApJ...887..220Y}. The variation of the negative magnetic fluxes in the red box (as shown in Figure 4(a)) is measured, as shown in Figure 4(e). The red dashed line represents the start time of the event. It is clear that the negative magnetic fluxes constantly decrease whether before or after the eruption onset. Only magnetic cancellation rate is different before and after the eruption onset. The magnetic cancellation rate is 6.8$\times$10$^{17}$ Mx hr$^{-1}$  before the eruption, and 3.3$\times$10$^{18}$ Mx hr$^{-1}$ after the eruption onset. It seems that the magnetic cancellation drove the eruption, and then the eruption speeded up the magnetic cancellation.

\subsection{The Eruption of the MF, the Related Coronal Jets, and the Formation of the FCS}

\begin{figure*}
	\includegraphics[width=0.9\textwidth]{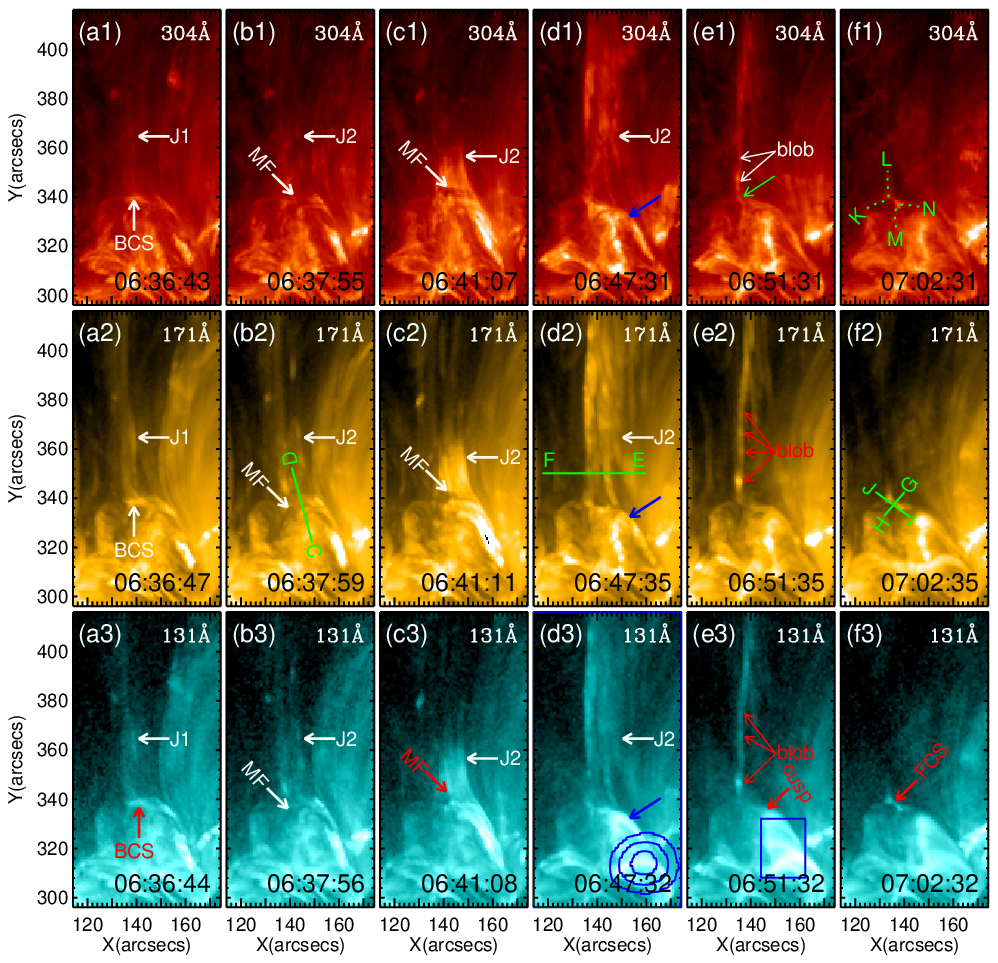}
    \caption{(a1 -- f1): SDO/AIA 304~{\AA}, (a2 -- f2): 171~{\AA}, and (a3 -- f3): 131~{\AA} images showing the jets and the formation of the flare current sheet (FCS). The lines from ``C'' to ``D'', from ``E'' to ``F'', from ``G'' to ``H'', from ``J'' to ``I'', from ``K'' to ``L'' and from ``M'' to ``N''  mark the slit position of the time-distance diagrams in Figures 9(a)-(f). The 131~{\AA} image with RHESSI hard X-ray sources superimposed is shown in panel (d3). The levels of the contours are at 50$\%$, 70$\%$, and 90$\%$ of the peak values in the 12–25 keV. The blue box in panel (e3) is used to calculate the changes of brightness during the microflare.}
    \label{fig:fig5}
\end{figure*}

Figure 5 presents the eruption of the MF, the induced coronal jets, and the formation of the FCS in SDO/AIA 304~{\AA}, 171~{\AA}, and 131~{\AA} wavelengths. As described above, the BCS formed at $\sim$06:35 UT, which could also be clearly detected in 304~{\AA} and 171~{\AA} wavelengths (see Figures 5(a1) and (a2)). It should be pointed out that a thin bright jet began to eject along the outer spine almost simultaneous with the appearance of the BCS. The jet is marked by ``J1'' in Figures 5(a1)-(a3). It looked very weak, and only existed for 2 minutes. The jet-like structure along outer spine mentioned above was part of J1. No cool materials were found in J1. As previously reported, jets generally exist rotation \citep{2011ApJ...728..103L,2012ApJ...745..164S,2013ApJ...769..134M,2023ApJ...945...96Y}. This jet was no exception. From animation 4, we found that J1 had a counterclockwise rotation when viewed from above. It's noted that the threads of J1 moved from left to right as it rose. J1 might be the result of the breakout reconnection between the magnetic field carrying the MF and the external open fields.  

\begin{figure*}
	\includegraphics[width=0.7\textwidth]{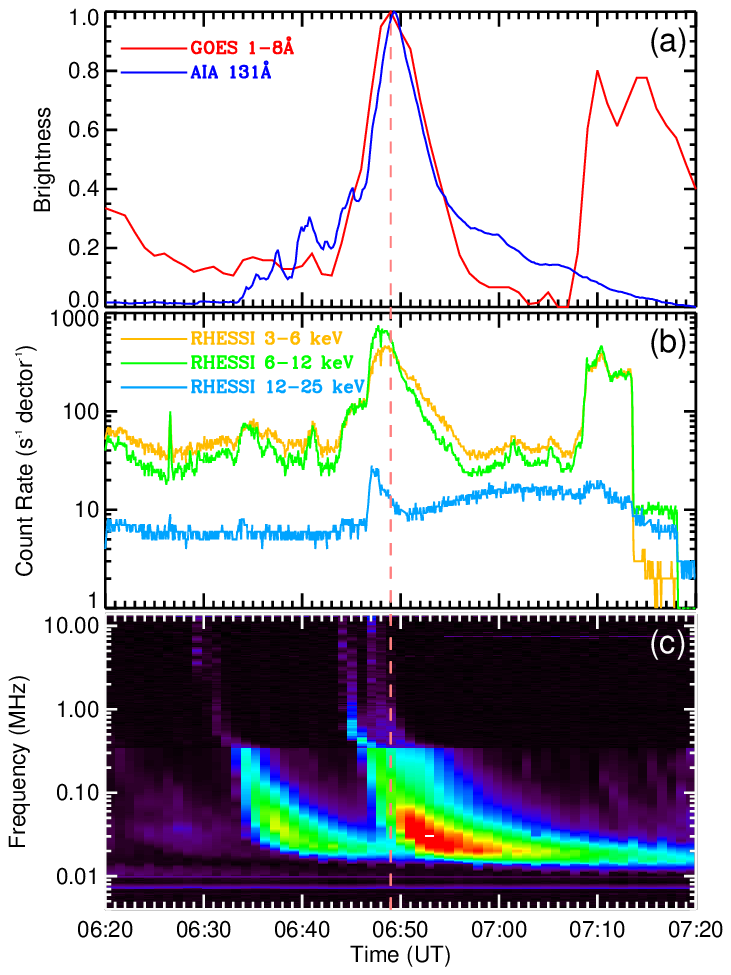}
    \caption{(a): GOES 1-8~{\AA} and SDO/AIA 131~{\AA} light curves, in units normalized to one. The AIA 131~{\AA} intensities are calculated in the microflare region marked by a blue rectangle in Figure 5(d3). (b): RHESSI 3--6 keV, 6--12 keV, and 12--25 keV light curves. (c): Radio dynamic spectra from the WAVES instrument on board the WIND spacecraft. The pink dashed line indicates the peak time of the microflare.}
    \label{fig:fig6}
\end{figure*}

During J1, the MF had been slowly rising. At $\sim$06:37 UT, the material of the MF started to be ejected as an another jet, marked as ``J2'' in Figure 5. J2 was obviously wider than J1. It is noting that the MF relatively rapidly erupted toward the northeast at $\sim$06:38 UT. It looked like a moving dark arch, with its northeast end always anchored. The eruption of a MF keeping one end of the MF still was ever observed by \citet{2019ApJ...879...74C}. For tracing the eruption of the MF, we made a time slice from 171~{\AA} images along a slice “CD” (as marked by a green line in Figure 5(b2)). As can be seen in Figure 9(a), the MF erupted at a velocity of 23.2$\pm$1.0 km s$^{-1}$. At $\sim$06:41 UT, the MF seemed to be constrained by the dome of the fan-spine structure. Soon afterwards, the whole MF erupted along the outer spine, and became the main part of J2 (see Figures 5(d1)-(d3)). And apparently, J2 consisted of bright (hot) and dark (cool) materials. J2 should be a blowout jet in the light of the definition of \citet{2010ApJ...720..757M}. From animation 4, it is found that J2 also had a counterclockwise rotation as viewed from above, which was the same with that of J1. J2 should possess negative helicity with left-handed chirality, which had the same chirality with the source MF. For tracing the rotational motion, we made a time–distance diagram along a slice “EF” (as marked by a green line in Figure 5(d2)) by using 171~{\AA} images. As can be seen in Figure 9(b), there existed three bright striped structures, suggesting J2's transverse rotating motion. The projection rotational velocity ranged from 27.2$\pm$3.0 km s$^{-1}$ to 39.9$\pm$2.0 km s$^{-1}$. This velocity might be an upper limit of the true rotation speed, because it combined the rotation velocity and axial expansion velocity.

\begin{figure*}
	\includegraphics[width=0.8\textwidth]{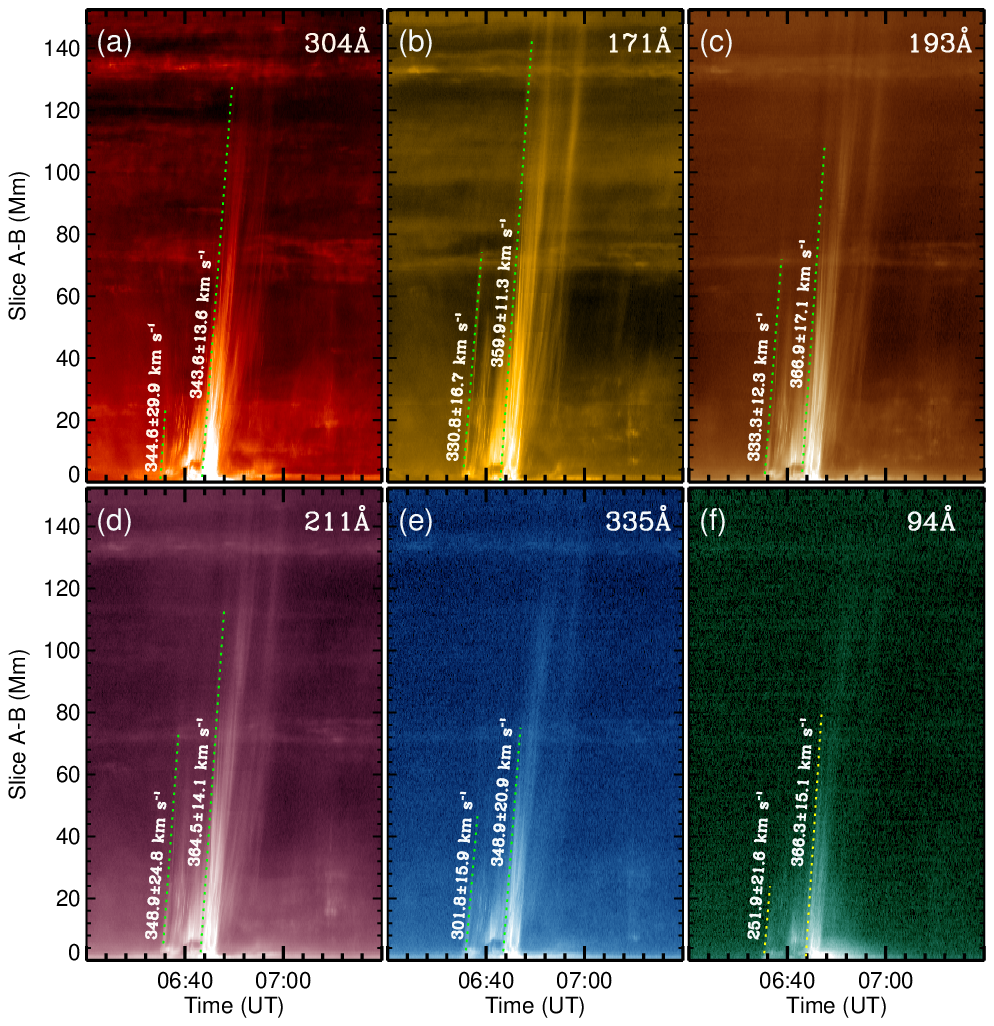}
    \caption{Time–distance diagrams along the line from ``A'' to ``B'' marked in Figure 1(b).}
    \label{fig:fig7}
\end{figure*}

The eruption of the MF was accompanied by a B9.5 class microflare (marked by the blue arrows in Figures 5(d1)-(d3)). Figures 6(a) and (b) show GOES soft X-ray 1-8~{\AA},  SDO/AIA 131~{\AA}, and RHESSI hard X-ray light curves of this flare. The GOES soft X-ray fluxes began to rise at about 06:42 UT, and reached its maximum at around 06:49 UT. The calculated SDO/AIA 131~{\AA} emission over the flare region has similar temporal evolution characteristics with the GOES soft X-ray emission. They reach their maximum at about the same time. The hard X-ray light curves seem to rise simultaneously with the GOES soft X-ray light curve, but they have earlier peak times at about 06:47 UT. The temporal difference between the peak times of the soft X-ray and hard X-ray light curves might indicate the Neupert effect of the microflare \citep{1968ApJ...153L..59N,2005ApJ...621..482V}. There was an X-ray source associated with this microflare. As shown in Figure 5(d3), the blue intensity contours of the hard X-ray emissions at 12-25 keV just laid on the bright and compact flare kernel. The radio dynamic spectra from WIND/WAVES is displayed in Figure 6(c). It is noted that a type III radio burst is associated with this microflare. The burst started at $\sim$06:47 UT, and rapidly drifted from more than 10MHz to near 0.1MHz. The start time of the burst is consistent with the peak time of the hard X-ray light curves. Type III radio bursts are generally produced by sub-relativistic electron beams propagate from the Sun to interplanetary space along open magnetic fields \citep{2014RAA....14..773R}. Therefore, there exist some open fields near the microflare site.

At $\sim$06:48 UT, an X-shaped structure appeared in 304~{\AA} and 171~{\AA} wavelengths (see the animation 4). It was gradually apparent at $\sim$06:51 UT (marked by the green arrow in Figure 5(e1)). The X-shaped structure consisted of one sheet-like structure and two bifurcated structures (see Figures 5(f1)-(f3)). The two bifurcated structures located at two end points of the sheet-like structure. The bifurcated structure was the cusp region connecting the reconnected magnetic field lines. Such structure resembled the reconnecting magnetic structure from two-dimensional magnetic reconnection \citep{1957JGR....62..509P}. The southwest end point of the sheet-like structure connected the post-flare loops after the eruption of MF, and the northeast end connected the outer spine and the dome field line. This connectivity was similar to the scenario of the FCS formation simulated by \citet{2018ApJ...852...98W}. The sheet-like structure was the FCS. It persisted about 34 minutes, and completely disappeared at $\sim$07:24 UT. With the formation of the X-shaped structure, a series of plasma blobs appeared in the outer spine, indicating that magnetic reconnection occurred (see Figures 5(e1)-(e3)). These plasma blobs in jets were ever observed by many simulations and observations, and their formation was considered to be the result of the tearing-mode instability \citep[e.g.,][]{2000A&A...360..715K,2014A&A...567A..11Z,2017ApJ...841...27N,2017ApJ...851...67S,2022FrASS...8..238C,2023MNRAS.518.2287M}. 

In order to investigate the kinematics features of coronal jets, we made time–distance diagrams by using 304~{\AA}, 171~{\AA}, 193~{\AA}, 211~{\AA}, 335~{\AA}, and 94~{\AA} images along a slit “AB” (Figures 7(a)– (f)). As can be seen, the jets appear as bright narrow-bands in the diagrams. To determine the velocities of the jets, a linear fitting was performed between the selected two points along the bright feature. We repeated the measurement 10 times, calculated the average value as the final velocity, and took the standard deviation as the error. It should be noted that only the main body velocities of the jets was measured. One can see that the jets had similar velocities in different wavelengths. It is found that J1 had a velocity range from 251.9$\pm$21.6 km s$^{-1}$ to 348.9$\pm$24.8 km s$^{-1}$, and J2 had a velocity range from 343.6$\pm$13.6 km s$^{-1}$ to 366.9$\pm$17.1 km s$^{-1}$. Notably, no jet material fell back to the solar surface, indicating that the jet material entirely ejected into large-scale loops or high corona. This coronal jet was not associated with a CME, but associated with a type III radio burst, indicating that there were non-thermal electrons escaping along open field lines.

\subsection{The Physical Characteristics of CSs}

\begin{figure*}
	\includegraphics[width=0.8\textwidth]{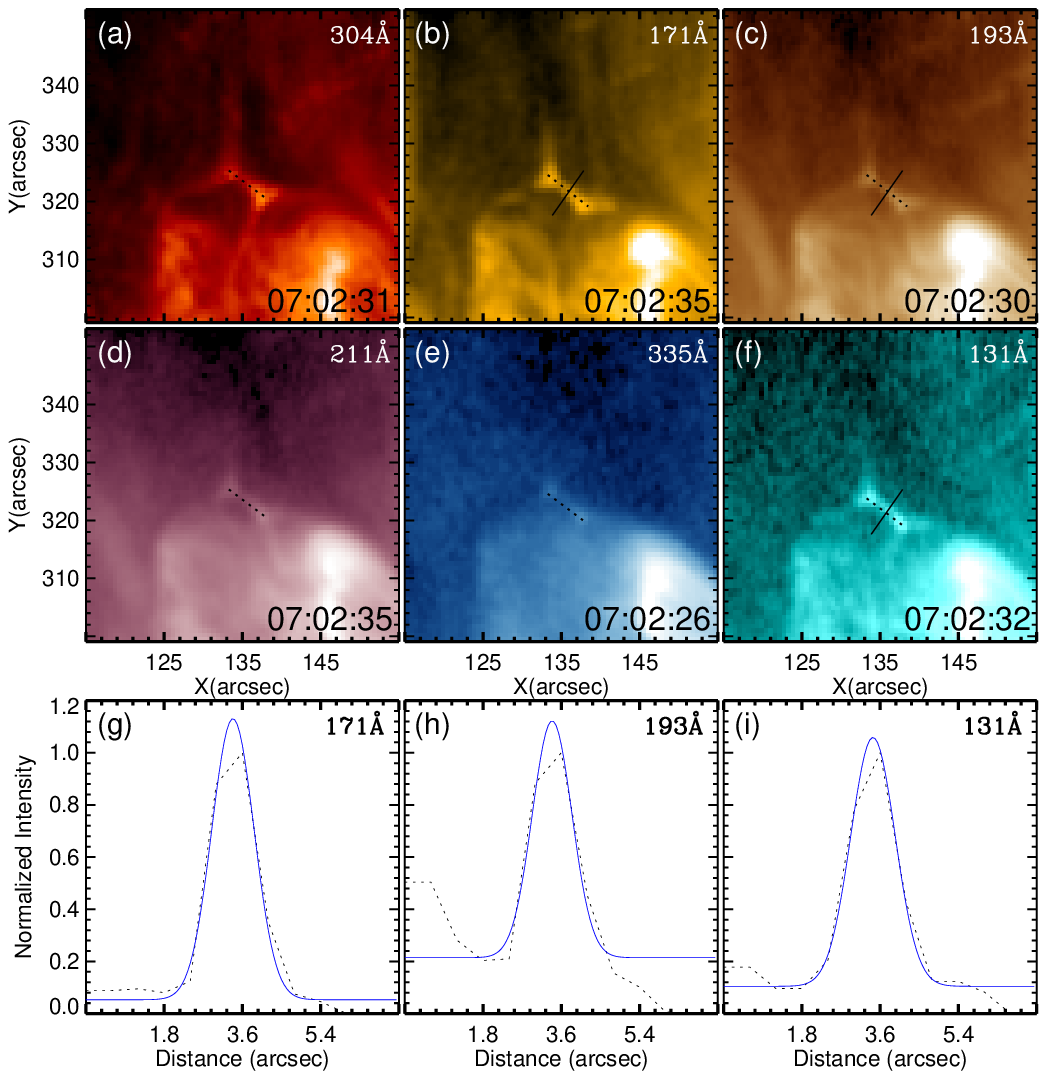}
    \caption{(a -- f): SDO/AIA 304~{\AA}, 171~{\AA}, 193~{\AA}, 211~{\AA}, 335~{\AA}, and 131~{\AA} images showing the FCS, respectively. The black dotted lines in panels (a -- f) are used to measure the length of the FCS. The black lines in panels (b), (c) and (f) denote the slit position at which we measure the width of FCS. (g -- i): normalized intensity profiles perpendicular to the FCS along the slit in panels (b), (c) and (f), with a Gaussian fit marked by the blue curve, respectively.}
    \label{fig:fig8}
\end{figure*}

Figure 8 presents the FCS in  304~{\AA}, 171~{\AA}, 193~{\AA}, 211~{\AA}, 335~{\AA}, and 131~{\AA} wavelengths. The black dotted lines in the panels are used to measure the length (l) of FCS. Its length was almost the same in different wavelengths, and was calculated to be from 4.35 to 4.93 Mm. The width (d) of FCS was estimated by the full width at half maxima (FWHM) of the brightness distribution in the direction perpendicular to the FCS. The width of FCS was also similar in different wavelengths, and was estimated to be 1.31 to 1.45 Mm. The estimates of the widths was biased by a projection effect, and it might be a lower limit of the true widths. The length and width of the FCS is obviously smaller than those calculated by \citet{2019ApJ...879...74C}. In a steady-state reconnection, the reconnection rate (M$_{r}$) is approximately equal to $d/l$ \citep{2000mare.book.....P,2016NatCo...711837X}. The reconnection rate M$_{r}$ is estimated to be from 0.266 to 0.333. As for the BCS observed above, we do not calculate its length and width because it might be incomplete due to the observation angle.

\begin{figure*}
	\includegraphics[width=0.8\textwidth]{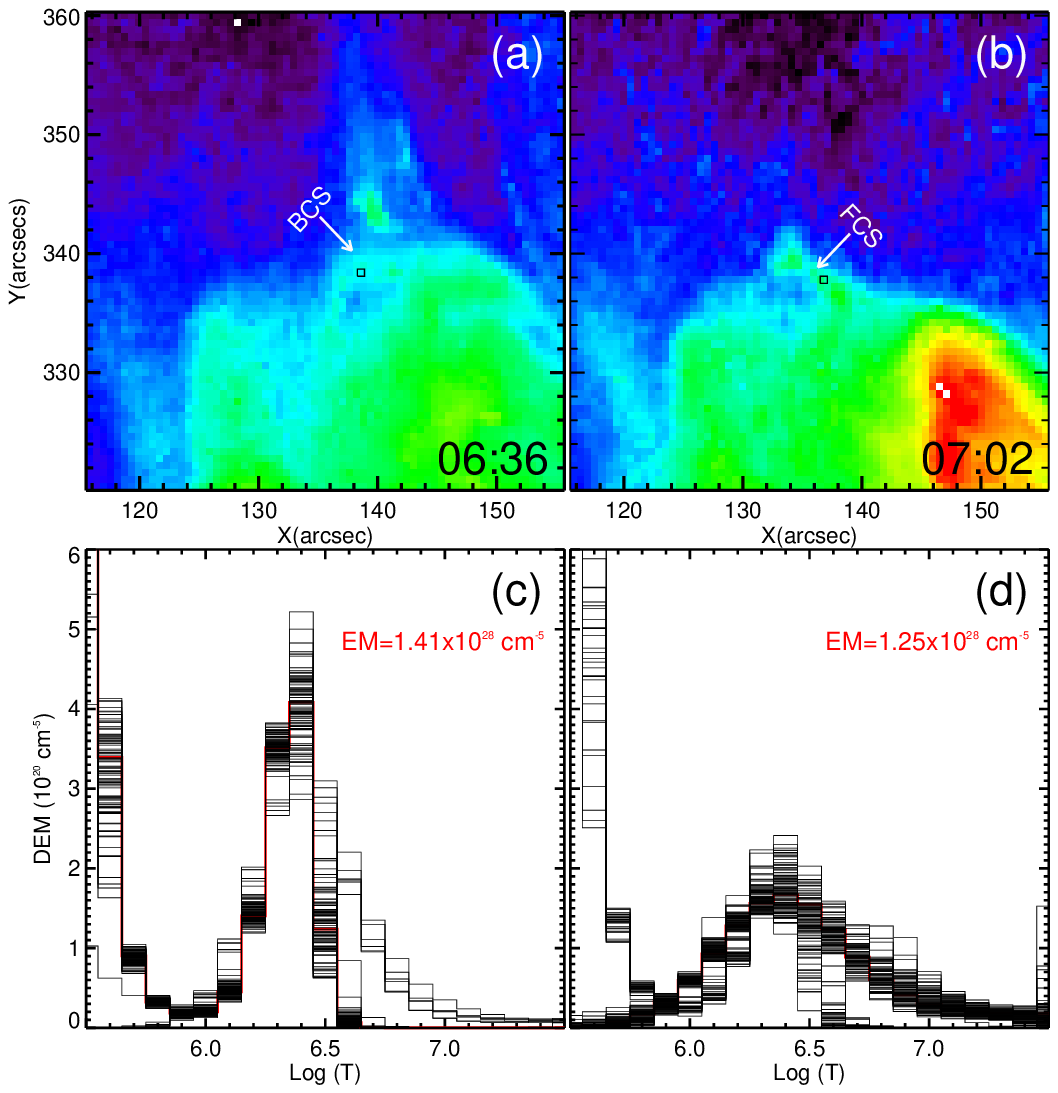}
    \caption{(a): The constructed emission measure image at 06:36 UT showing the BCS. (b): The constructed emission measure image at 07:02 UT showing the FCS. (c): The averaged DEM distribution of the BCS outlined by the black box in panel (a). (d): The averaged DEM distribution of the FCS outlined by the black box in panel (b). The black curves represent 100 Monte Carlo simulations, and the red curve denotes the best-fit DEM curve.}
    \label{fig:fig9}
\end{figure*}

The emission measure (EM) maps of the BCS (06:36 UT) and the FCS (07:02 UT) were present in Figure 9. The BCS and the FCS were identified as enhanced emission regions in the EM maps. The average DEM distributions of the BCS and the FCS were shown in Figures 9(c)–(d), respectively. It can be seen that the average DEM distribution of the BCS ranged from log $\emph{T}$ = 6.0 (1MK) to log $\emph{T}$ = 7.0 (10 MK), and the average DEM distribution of the FCS ranged from log $\emph{T}$ = 5.8 (0.6 MK) to log $\emph{T}$ = 7.2 (15.8 MK). The DEM distributions of the BCS and FCS were wide, indicating that the BCS and the FCS were multi-thermal structures. It is worth noting that the DEM distributions of the BCS and the FCS both had unimodal structures. The peak temperature of BCS and FCS both were log $\emph{T}$ = 6.4 (2.5 MK). 

\subsection{Plasma Flow}

In magnetic reconnection theory, plasma blobs might ubiquitously exist in a CS due to the tearing-mode instability \citep{2005ApJ...622.1251L}. For tracing the reconnection inflows, we made a time slice using 171~{\AA} images along a slice “G-H” (as marked by a green line in Figure 5(f2)). Unfortunately, no obvious reconnection inflows were observed. However, intermittent plasma blobs were detected in the CS (as marked by red arrows in Figure 10(c)). To study the motion of the plasma blobs in the CS, we made a time slice from 171~{\AA} images along a slice “I-J” (as marked by a green line in Figure 5(f2)). It can be seen from Figure 10(d) that there were many bright strips, suggesting the rapid movement of the plasma blobs. Only part of strips were traced, and the velocities of the plasma blobs ranged from 46.1$\pm$3.2 km s$^{-1}$ to 93.2$\pm$9.6 km s$^{-1}$. 

\begin{figure*}
	\includegraphics[width=0.8\textwidth]{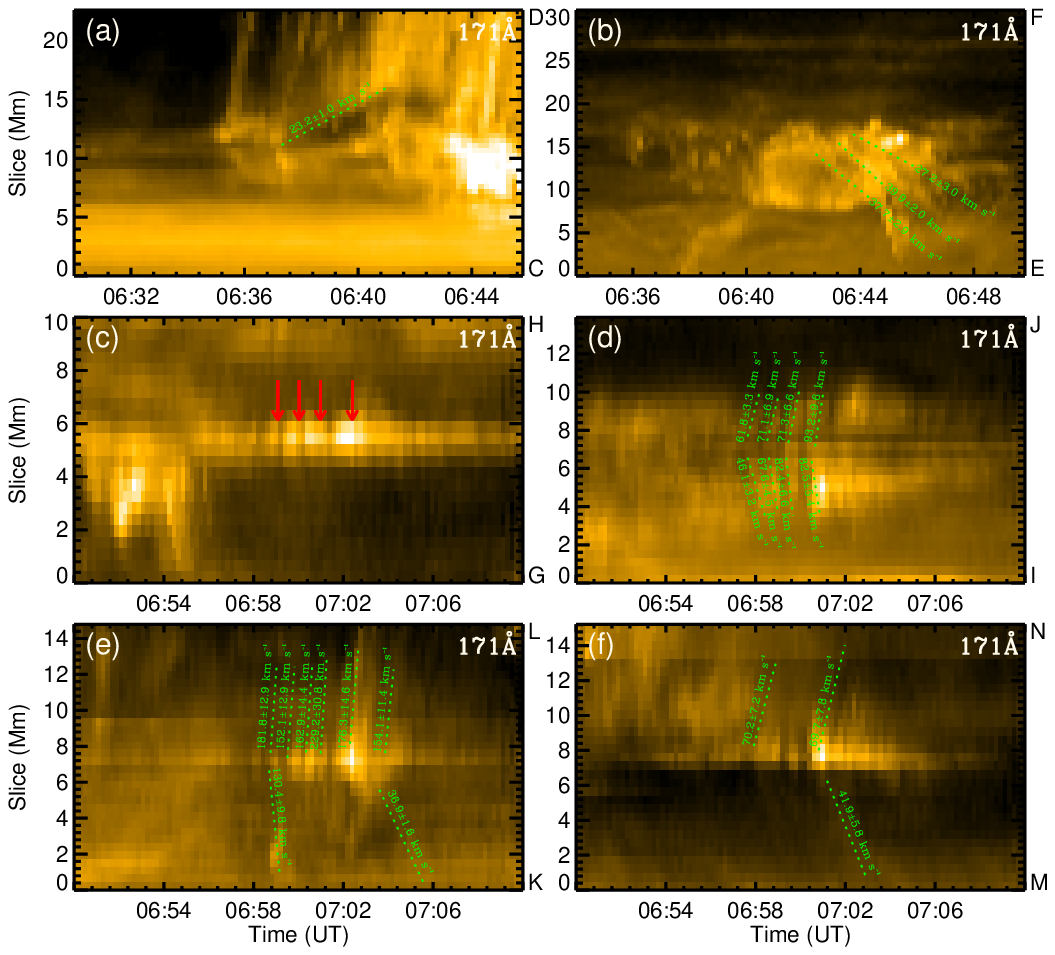}
    \caption{(a -- f): Time–distance diagrams along the lines from ``C'' to ``D'' marked in Figure 5(b2), from ``E'' to ``F'' marked in Figure 5(d2), from ``G'' to ``H'' marked in Figure 5(f2), from ``I'' to ``J'' marked in Figure 5(f2), from ``K'' to ``L'' marked in Figure 5(f1), and from ``M'' to ``N'' marked in Figure 5(f1).}
    \label{fig:fig10}
\end{figure*}

For investigating the physical properties of plasma blobs along the reconnecting lines, we made time slices using 171~{\AA} images along slices “K-L” and “M-N” (as marked by green lines in Figure 5(f1)). The results are presented in Figures 10(e) and (f). The plasma blobs were recognized as bright strips in the time–distance diagrams. Their existence suggests the tearing-mode instability occurring in the BCS during the magnetic reconnection process. It is noted that the plasma blobs did not symmetrically appear in the reconnecting lines. It seems that the plasma blobs intensively appeared in one bifurcation of the reconnecting lines. The velocities of the plasma blobs along the outer spine were fastest, and ranged from 152.1$\pm$12.9 km s$^{-1}$ to 229.2$\pm$12.9 km s$^{-1}$. The velocities of the plasma blobs along the dome ranged from 36.9$\pm$1.6 km s$^{-1}$ to 130.4$\pm$9.8 km s$^{-1}$. The velocities of the plasma blobs along the post-flare loops were ranged from 41.9$\pm$5.8 km s$^{-1}$ to 70.2$\pm$7.2 km s$^{-1}$.

\section{Conclusions and Discussions}

CS is one of the vital physical features of magnetic reconnection. It is extensively observed in large-scale magnetic reconnection events, but still a challenge to be 
observed in small-scale magnetic reconnection events, especially in coronal jets,  because its observation is affected by not only the respond of observational instruments, but also the observed angle. In this paper, it is lucky for us that two CSs (a BCS and a FCS) were simultaneously observed in one jet event with high resolution data from the SDO, the NVST, the RHESSI, the Wind, and the GOES. In the source region of coronal jets, one MF with negative helicity was observed to lay on the PIL, which was confined in a fan-spine magnetic structure. The MF was first activated by flux cancellation under it, and then it slowly rose. When the magnetic fields wrapping the MF squeezed to the antidirectional external open fields, a BCS formed, and a thin bright coronal jet was observed to move along the outer spine of the fan-spine magnetic structure. Simultaneously, two bidirectional jet-like structures were observed to eject along the outer and inner spine. The jet-like structure along the outer spine was part of the observed coronal jet. As the MF rapidly erupted, a blowout jet was observed. It had a counterclockwise rotation, whose helicity was consistent with that of the MF. Behind the erupting MF, a FCS was formed. One end of the FCS connected the post-flare loops of the erupted MF. This process accords with the standard flare model. The peak temperature of the BCS was calculated to be 2.5 MK. The length, width and peak temperature of the FCS was calculated to be 4.35-4.93 Mm, 1.31-1.45 Mm, and 2.5 MK, respectively. The magnetic reconnection rate associated with the FCS was estimated to be from 0.266 to 0.333. This event was associated with a type III radio burst in the frequency from 0.1MHz to 10MHz, revealing its influence on the interplanetary space. These observations support the scenario of the breakout model of coronal jets, and flux cancellation might be the driver of this event.

The evolutionary process of this event is very similar to the breakout model for solar coronal jets with filaments simulated by \citet{2017Natur.544..452W,2018ApJ...852...98W}. First, both have similar initial magnetic field structure. The initial setting of the simulation is a potential field, in which a compact bipolar structure is embedded in a uniform inclined background field. The setup produces a confining field beneath the three dimensional null points. Such structure is similar to the pre-eruption fan-spine structure of this event. Second, both have similar reconnection process. In the simulation, the BCS creates when the restraining field of the sheared field expands upwards towards the null point. The BCS can produce a slow, spiry plasma outflow. Massive energy release occurs only when the magnetic field containing the MF reaches the breakout sheet and reconnects with the open field. The rapid reconnection between the magnetic field containing the MF and the background open field launches an untwisting jet, and initiates flare reconnection, which induces the formation of a FCS and flare loops. Similar to the simulation, a BCS, a helical blowout jet, a FCS, and a micro-flare were observed in this event. Different from the simulation, the BCS not only produced a thin bright jet, but also produced bidirectional jet-like structures. In this simulation, the MF experienced slow-rise phase and fast-rise phase. However, only fast-rise phase of the MF in this event was measured, perhaps due to the restraining field of the MF was near the null point. On the whole, this event is consistent with the jet breakout model of \citet{2017Natur.544..452W} and \citet{2018ApJ...852...98W}. 

In this event, the FCS formed during the B9.5 class microflare, and the southwest end point of the FCS connected the post-flare loops. However, it is not a typical FCS. It persisted a long time, and still remained after the microflare stopped. It even became clearer after the microflare. We surmise that it might be part of the three-dimensional CS, and make some contributions to the microflare, but not all.  Although the microflare is over, local magnetic reconnection might still continue. But of course, there is an another possibility. The CS after the microflare might be the slowly reconnecting current layer at the null point in the relaxation phase simulated by \citet{2018ApJ...852...98W}. The current layer means the occurrence of the interchange reconnection, which stops when the remained magnetic configuration relaxes toward a new equilibrium. Such a long-lasting FCS was ever studied by \citet{2020ApJ...900...17Y}  in a large flare. Their studied FCS is much larger than the one studied in this event, whose length extends to greater than 10 R$_\odot$. They believed that their observational results accords with the standard Carmichael-Sturrock-Hirayama-Kopp-Pneuman (CSHKP) flare model for gradual phase. In their event, sporadic magnetic reconnections are found to occur at the magnetic null point in the FCS, which result in bidirectional plasma outflows. Each arrival of sunward outflows at the cusp-shaped loop top is along with an impulsive microwave and X-ray burst, inducing plasma heating and particle acceleration in the post-flare arcades. Similar to their observations, bidirectional plasma outflows in the FCS were also clearly observed in this event (see Figure 10(d)). However, no impulsive microwave and X-ray burst were found at the top of the cusp-shaped flare arcade in this event. It is suspected that the plasma outflows perhaps carried not enough energy to establish a fast-mode termination shock in the cusp region, which could accelerate energetic electrons and heat plasma. The existence of the termination shock was proposed by \citet{2015Sci...350.1238C} in a long-duration flare. In their work, the instantaneous spatial distribution of the radio spikes at different frequencies forms a narrow surface at the loop-top region, whose location and morphology closely resemble those of a termination shock. In short, the small-scale and large scale FCSs have many similarities due to similar magnetic reconnection mechanisms, but some differences due to different scales.

By studying the role of small-scale filament eruptions in the generation of X-ray jets in coronal holes, \citet{2015Natur.523..437S} proposed a two-step reconnection model. Internal reconnection firstly happens in the restraining field of the MF, leading to the eruption of the MF. External reconnection then occurs between the erupting MF field and the ambient open field. Unlike their observations, external reconnection first occurred between the magnetic field carrying the MF and the open field of the fan-spine structure, and then internal reconnection occurred between the distended legs of the MF in this event. The two-step reconnection processes in coronal jet was ever reported by other researchers. \citet{2018ApJ...854..155K} reported that the eruption process of an equatorial coronal-hole jet accords with the magnetic breakout model. The slow external magnetic reconnection is found to take place between the overlying closed flux of the flux rope with the external field near the dome, producing quasi-periodic mass flows along the spine. Internal magnetic reconnection then occurs beneath the rising flux rope, producing a thin, bright features called FCS. Recently, \citet{2023ApJ...942...86Y} present the observational signatures of external reconnection and  internal reconnection in studying a coronal hole jet. 

The studied jet eruption with a B9.5 class microflare was related to a type III radio burst and a hard X-ray source, which can be diagnostic tools of energetic electrons in flares. If the energetic electron beams interact with the plasma in the high corona or interplanetary space, Langmuir wave will be produced and then decay into electromagnetic waves, which can be observed as type III radio bursts. The type III radio burst is usually interpreted as a signature of propagating beams of non-thermal electrons escaping from the Sun into the interplanetary space \citep{2014RAA....14..773R}. This demonstrates that such a small event indeed impacted the space weather. Previous researches demonstrated that the interchange reconnection occurs between the closed magnetic fluxes of the source region and the ambient open magnetic fluxes if a coronal jet is along with a type III radio burst \citep[e.g.,][]{2008ApJ...675L.125N,2011ApJ...742...82K,2017ApJ...835...35H,2018ApJ...866...62C}. In this event, the non-thermal electrons were first produced by the microflare at the jet base, and then they were carried away by the open field through the interchange reconnection between the magnetic field containing the MFs and the open field. If the energetic electron beams interact with the plasma in the low corona and chromosphere, bremsstrahlung hard X-ray emission will be produced \citep{2020ApJ...889..183M}. Some prior studies have shown that a hard X-ray source can be observed in the jet base if the jet is along with a type III radio burst \citep[e.g.,][]{2011ApJ...742...82K,2017ApJ...835...35H,2018ApJ...866...62C}.

As mentioned above, these small-scale CSs in coronal jets are actually hard to be observed due to the sensitivities of the observed instruments and the observed angles. Their physical parameters are important for the coronal jet models. Breakout reconnection preceding a jet near an AR was studied by \citet{2019ApJ...874..146H}, and they found the BCS had a length less than 3 $''$ ($\sim$2.2 Mm), a width less than 1 $''$ ($\sim$0.7 Mm), and a peak temperature of 1.6 MK. The peak temperature of their BCS is smaller than that of our event (2.5 MK). \citet{2019ApJ...879...74C} measured the length (10.8 and 7.51 Mm), the width (1.86 and 3.4 Mm), and the temperature (1.81 and 1.77 MK) of FCSs of two coronal jets near an AR. Our studied FCS is obviously shorter and narrower than theirs. \citet{2018ApJ...854..155K} detected multiple blobs with projected speeds of 135 and 55 km s$^{-1}$ in the bright, inverted-V-shaped structure below the flux rope. The velocities of plasma blobs along the reconnected field lines were measured to be from 109 to 178 km s$^{-1}$ \citep{2019ApJ...879...74C}. The velocities of plasma blobs observed in this event (from 39.6$\pm$1.6 km s$^{-1}$ to 229.2$\pm$12.9 km s$^{-1}$) had the same order of magnitudes with previous observations. 

Whether the magnetic flux emergence or cancellation driving the jet event has always been controversial. Some researchers present evidence that coronal jets appear where magnetic flux cancellation is occurring \citep{2011ApJ...738L..20H,2012A&A...548A..62H,2017ApJ...844..131P,2017ApJ...840...54C,2018ApJ...853..189P,2018ApJ...864...68S,2019ApJ...881..132D,2023ApJ...942...86Y}. Some studies found that magnetic flux emergence might play a key role in producing coronal jets \citep{2007A&A...469..331J,2015Ap&SS.359...44L,2015SoPh..290..753S}. In ARs, situations might be more complicated. It is difficult to separate emergence from cancellation because both occur frequently during some times of the region's life \citep{2017ApJ...844...28S,2023ApJ...945...96Y,2017Ap&SS.362...10J}. For this event, although it occurs at the edge of an AR, the jet-driving mechanism is no doubt. Magnetic flux cancellation under the MF drives the entire event. Photospheric magnetic flux cancellation might be associated with internal magnetic reconnection of the MF. However, we can not provide evidence that how the magnetic reconnection occur. But anyway, the photospheric magnetic flux cancellation did lead to the activation and slow rise of the MF. The slow rise of the MF further leads to the external reconnection occurring between the magnetic field carrying the MF and the open field of the fan-spine structure, the internal reconnection occurring between the distended legs of the MF, and two coronal jets.

\section*{Acknowledgements}

The authors are grateful to the anonymous referee for his/her constructive suggestions and comments, which led to the great improvement of this manuscript. The authors are indebted to the NVST and SDO science teams for providing the data. The authors would like to thank Prof. Yuandeng Shen, Dr. Junchao Hong, Dr. Yi Bi and Dr. Hechao Chen for their useful discussions. The authors thank Kexing Li for helping to download the data. This work is supported by the Strategic Priority Research Program of the Chinese Academy of Sciences, Grant No. XDB0560000, the National Key R\&D Program of China (2019YFA0405000, 2021YFA1600502, and 2022YFF0503800), , Yunnan Key Laboratory of the Solar physics and Space Science (YNSPCC202212), National Natural Science of China (12325303, 12173084, 11973084, 12203097, 12003064, 12373065, and 12273060), Yunnan Science Foundation of China (202101AT070032, 202201AT070194, 202301AT070349, and 202301AT070347), the Key Laboratory of Solar Activity of CAS (KLSA202101), Youth Innovation Promotion Association, CAS (Nos 2019061 and 2023063) , CAS “Light of West China” Program, and Young Elite Scientists Sponsorship Program by Yunnan Association for Science and Technology.

\section*{Data Availability}
The data underlying this article will be shared on reasonable request to the corresponding author.

\appendix


\bsp	
\label{lastpage}
\end{document}